\documentclass[12pt,preprint]{aastex}
\def\kms{km s${}^{-1}$}
\def\ab{$\sim$}

\def\p{$\pm$}
\def\etal{{et~al.}}
\def\deg{$^\circ$}
\def\m348{Mrk~348}
\def\solmass{M$_\odot$}
\def\HI{H\kern0.1em{\sc i}}
\def\h2o{H$_2$O}

\font\caps=cmcsc10 scaled 1200

\shorttitle{Flaring Maser in \m348}
\shortauthors{Peck \etal}
\received{2002 November 24}
\begin{document}

\title{The Flaring \h2o Megamaser and Compact Radio Source in \m348} 

\author{A. B. Peck\altaffilmark{1,2}, C. Henkel\altaffilmark{1},
J. S. Ulvestad\altaffilmark{3}, A. Brunthaler\altaffilmark{1},
H. Falcke\altaffilmark{1}, M. Elitzur\altaffilmark{4},
K. M. Menten\altaffilmark{1}, J. F. Gallimore\altaffilmark{5}}

\altaffiltext{1} {Max-Planck-Institut f\"ur Radioastronomie, Auf dem
H\"ugel 69, D-53121 Bonn, Germany}
\altaffiltext{2}{current: Harvard-Smithsonian Center for Astrophysics, SAO/SMA
Project, P.O. Box 824, Hilo, HI 96721, USA, apeck@cfa.harvard.edu}
\altaffiltext{3}{National Radio Astronomy Observatory, P.O. Box O,
Socorro, NM 87801} \altaffiltext{4}{Dept. of Physics and Astronomy,
University of Kentucky, Lexington, KY 40506-0055}
\altaffiltext{5}{Dept. of Physics, Bucknell University, Lewisburg, PA
17837}

\setcounter{footnote}{0}

\begin{abstract}
We report on single-dish monitoring and extremely high angular
resolution observations of the flaring H$_2$O megamaser in the Seyfert 2
galaxy \m348.  The H$_2$O line is redshifted by $\sim$130~km s$^{-1}$
with respect to the systemic velocity, is very broad, with a FWHM of
130~km s$^{-1}$, and has no detectable high velocity components within
1500~km s$^{-1}$ on either side of the strong line.  Monitoring
observations made with the Effelsberg 100m telescope show that the
maser varies significantly on timescales as short as one day and that
the integrated line flux is loosely correlated with the continuum
flux.  VLBA observations indicate that the maser emission arises
entirely from a region less than 0.25~pc in extent, located toward a
continuum component thought to be associated with the receding jet.
We also report on integrated flux monitoring with the VLA between 1.4
and 43~GHz, and VLBA continuum observations of the milliarcsecond
scale jets at 1.7, 8, 15, and 22~GHz.  These observations have allowed
us to tentatively pinpoint the location of the core, and also show
the ejection of a new jet component during the current radio
``flare.''

\end{abstract}
\keywords{masers -- galaxies:active -- galaxies:jets -- radio
lines:galaxies -- galaxies:Seyferts -- galaxies:individual
(\m348,NGC~262)}

%\twocolumn
%\clearpage
\section{Introduction}

\subsection{\h2o\ Megamasers}

\h2o\ megamasers are best known as a means to probe the accretion
disks in Seyfert galaxies. In the most famous source, NGC~4258, a
thin, slightly warped, nearly edge-on disk orbits in Keplerian
rotation around a central mass of 4$\times$10$^{7}$ \solmass\
(e.g. Greenhill \etal\ 1995; Miyoshi \etal\ 1995, Herrnstein \etal\
1999).  VLBI studies have been used to determine the size and shape of
this warped molecular disk as traced by the maser spots.  A few other
sources show evidence of a toroidal structure, but the distribution of
maser spots is not as well understood.  For example, VLBI observations
of megamasers seen in NGC~3079 indicate that the structure of the
maser components is consistent with a parsec-scale disk, but also
exhibits a significant non-rotational component (Trotter \etal\ 1998).
NGC~3079 also exhibits \HI\ absorption toward the core (Sawada-Satoh
\etal\ 2000; Baan \& Irwin 1995), which in some cases can be a tracer
of a circumnuclear torus ({\it e.~g.} Peck \& Taylor 2001).  The
NGC~4945 maser (Greenhill, Moran \& Herrnstein 1997) is tentatively
interpreted to consist of a parsec-scale disk, but shows some
deviations in both position and velocity of the maser components from
the Keplerian model.

There is evidence, however, for a distinct class of H$_2$O
megamaser. In these sources the amplified emission is the result of an
interaction between the radio jet and an encroaching molecular cloud,
rather than occurring in a circumnuclear disk.  The only known sources
in this class were NGC~1068 (Gallimore \etal\ 1996) and the Circinus
galaxy (Greenhill \etal\ 2001), which appear to have both a
circumnuclear disk and maser emission arising along the edges of an
ionization cone or outflow, and NGC~1052, in which the masers appear
to arise along the jet and have a full width at half maximum (FWHM)
$\sim$90 \kms\ (Claussen \etal\ 1998).  We have recently identified
the fourth such source, \m348, a Seyfert 2 galaxy with a low
inclination angle and an exceptionally bright and highly variable
nuclear radio source (Falcke \etal\ 2000).

\subsection{\m348}

\m348 (NGC~262) is a well-studied galaxy which exhibits interesting
morphology at a wide range of wavelengths. This galaxy has been
classified as S0/a (deVaucouleurs, deVaucouleurs \& Corwin 1976),
although recent studies by Ant\'on \etal\ (2002) indicate that it is
more likely between Sa+ and Sb+.  It exhibits a large \HI\ halo
(Morris \& Wannier 1980) which may have been produced by an
interaction with a companion galaxy (NGC~266, at an angular distance
of 23$\arcmin$; Simkin \etal\ 1987).
%%The recessional velocity of \m348\ has been measured to be
%%c$z=$4507\p4 \kms\ based on nuclear \HI\ emission (Bottinelli \etal\
%%1990).  Other values ranging from c$z=$4197 \kms\ (optical emission
%%lines; Cruz-Gonzales \etal\ 1994) to c$z=$4660 \kms\ (optical emission
%%lines; Koski 1976) and c$z=$4800 \kms\ ([OIII] emission lines; Feldman
%%\etal\ 1982) can be found in the literature, as well as a number of
%%\HI\ emission measurements between 4490 and 4540 \kms\ ({\it e.~g.}
%%Richter \& Huchtmeier 1991; Mirabel \& Wilson 1984; Heckman, Balick \&
%%Sullivan 1978).  
A variety of measurements of recessional velocity, both optical and
radio, exist in the literature ({\it e.g.} Cruz-Gonzales \etal\ 1994,
Richter \& Huchtmeier 1991, Bottinelli \etal\ 1990, Mirabel \& Wilson
1984, Feldman \etal\ 1982, Heckman, Balick \& Sullivan 1978, Koski
1976).  It is possible that some of these measurements might be
affected by the gas in the halo.  Because the redshift of optical
emission lines can be affected by local dynamics and stellar outflows
in the galaxy, we adopt the recent central \HI\ value of Bottinelli
\etal\ (1990; c$z_{\rm hel}=$ 4507 \kms) as the most likely
estimate of the systemic velocity.  This corresponds to a distance of
62.5 Mpc, assuming H$_{\rm 0}$=75 \kms\ Mpc$^{-1}$.

VLBI observations of \m348\ at 1.417~GHz by Neff \& de Bruyn (1983)
show a three component radio source with a total extent of \ab180 mas
(\ab50 pc).  Later observations using Global VLBI at 4.8~GHz (Roy
\etal\ 1999) show two faint continuum hotspots on either side of a
much brighter central peak.  This central component is thought to be
the core, and the separation between the outer hotspots and the core
is around 50 mas to the north, and 120 mas to the south.  The relative
intensities of the two outer components are quite similar, indicating
that relativistic beaming effects are probably minimal and the jet
axis should be close to the plane of the sky.  Ground-based
observations (Simpson \etal\ 1996) show evidence of an ionization cone
with a half-angle of \ab45\deg, which also suggests a jet axis fairly
close to the plane of the sky.  VLBA images at higher frequencies
(Ulvestad \etal\ 1999) reveal a small-scale double continuum source,
the axis of which is aligned with the optical [OIII] emission (Capetti
\etal\ 1996), with a position angle of $-$16\deg.  Although a broad
polarized H$\alpha$ line with FWHM \ab7400 \kms\ (Miller \& Goodrich
1990) and a strongly absorbed hard x-ray source having $N_{\rm
H}=$10$^{23.1}$ cm$^{-2}$ (Warwick \etal\ 1989) argue for the presence
of an obscured nucleus, many attempts to detect the expected obscuring
torus at radio wavelengths have not been successful.  Gallimore \etal\
(1999) searched for \HI\ in absorption with the VLA in 1992, but did
not detect any absorption toward the radio source with a 3$\sigma$
upper limit of 4~mJy, or $\tau$\ab0.014.  Taniguchi \etal\ (1990) did
not detect any $^{12}$CO(J=1$-$0) emission or absorption using the
Nobeyama Radio Observatory 45-m telescope in 1989, although their
upper limit of 10.4 K \kms\ for this source was fairly high.
Barvainis \& Lonsdale (1998) reported that there was no evidence for
free-free absorption in the spectral energy distributions of the core
component in \m348 between 1.4 and 15~GHz.  Hubble Space Telescope
(Falcke, Wilson \& Simpson 1998) observations show evidence of a dust
lane crossing the nucleus.

%%The compact radio source in \m348\ is unique among Seyferts in that it
%%is very bright and extremely variable.  The continuum flux during the
%%single-dish monitoring program varied from 640 to 860~mJy.  The high
%%resolution observations of the maser line presented here were made
%%during a local minimum, when the total continuum flux density at 22~GHz was $%%\sim$0.6 Jy.

In this paper, we present the results of Effelsberg 100-m telescope
monitoring of the evolution of the \h2o\ line flux and 22~GHz total
continuum flux density ($\S$3.1), VLA monitoring of the integrated
continuum flux density at various frequencies ($\S$3.2), observations
of the milliarsecond scale jet morphology during the course of the
continuum flare ($\S$3.3) and high spatial and spectral resolution
VLBA imaging of the megamaser emission at one epoch in June 2000
($\S$3.4).  Details of the methods used for all of these experiments
are given in $\S$2.  In $\S$4, we present discussions of the physical
environment in the central few parsecs in this galaxy, and the conditions
which give rise to megamaser emission.

\section{Observations and Analysis}

\subsection{Effelsberg observations}

With the Effelsberg 100-m telescope\footnote {The 100-m telescope at
Effelsberg is operated by the Max-Planck-Institut f\"ur
Radioastronomie (MPIfR) on behalf of the Max-Planck-Gesellschaft
(MPG).}, data were taken at the end of 1997 and early 1998 in a
position switching mode with a 22 GHz maser receiver and a backend
consisting of 1024 channels and a bandwidth of 50~MHz.  The recorded
system temperature was $T_{\rm sys}$ $\sim$ 75~K on a $T_{\rm A}^{*}$
(antenna temperature corrected for atmospheric attenuation)
temperature scale. Between March 2000 and August 2001 we used a new
dual channel 22 GHz HEMT receiver in a dual beam switching mode with a
beam throw of 2\arcmin\ and switching frequency of 1\,Hz (for details,
see Falcke \etal\ 2000). For the measurements with the HEMT receiver,
pointing could be checked on \m348\ itself.  Amplitude calibration was
obtained using NGC~7027 (5.8 Jy; see Baars \etal\ 1977, Ott \etal\
1994) as a primary calibrator and adopting the 22~GHz gain curve given
by Gallimore \etal\ (2001).  Since \m348\ itself could be used as a
pointing source, pointing errors do not significantly affect
calibration uncertainties so that we conservatively estimate that our
flux densities are accurate to \p10\% or better.

\subsection{VLA monitoring}
The NRAO\footnote{The National Radio Astronomy Observatory is a
facility of the National Science Foundation operated under cooperative
agreement by Associated Universities, Inc.} Very Large Array (VLA)
observations were made at six frequencies ranging from 1.4~GHz to
43~GHz. We observed \m348 on 8 dates between November 1998 and July
2001 in various configurations.  The source 3C48 was used as the
primary flux density calibrator, and \m348 was self-calibrated and
imaged with the Astronomical Image Processing System (AIPS; van
Moorsel, Kemball, \& Greisen 1996).  The VLA images are not shown as
the radio source was unresolved in all cases; only the integrated flux
densities are reported here.
                                                  
\subsection{VLBA Continuum Observations}

The NRAO Very Long Baseline Array (VLBA) was used to image \m348 in
the continuum at three epochs: 1997.10, 1998.75, and 2000.00.
Observing frequencies ranged from 1.7 through 22.2~GHz, with three
different frequency bands used at each epoch.  The 15~GHz observations
from the first two epochs have been described previously by Ulvestad
\etal\ (1999, hereafter U99), while the remainder of the observations
are reported for the first time here.  In all instances, the VLBA
observations utilized four 8-MHz IF channels with two-bit sampling of
the data, resulting in total data rates of 128~Mbit~s$^{-1}$.  At the
first two epochs, single polarizations were observed at each
frequency, while dual polarizations were used at the third epoch.
Between 1 and 3 hours was spent integrating on \m348 in each of the
frequency bands observed at each epoch.  Details of the observations
are summarized in Table~\ref{tab:vlba-cont}.

All data calibration was carried out in AIPS.  The amplitudes of the
VLBA data were calibrated by means of the standard antenna gain files
maintained by VLBA staff as well as system temperatures measured every
1--2 minutes during the observations.  This calibration is estimated
to be accurate to \p5\% at all frequencies through 15~GHz, and \p10\%
at 22~GHz due to the increased (variable) impact of atmospheric
emission.  At the first epoch, the antennas at North Liberty, Iowa and
Hancock, New Hampshire were not used due to snow and ice in the
dishes, while the Pie Town, New Mexico antenna failed during the
second epoch.  Low-elevation data (typically below 10\deg) were
discarded at the higher frequencies; other data editing was minor.
Residual clock errors were removed by means of observations of a
strong calibrator source or the application of the pulse-calibration
tones inserted at the VLBA front ends.  Residual delay, rate, and
phase errors were derived in most cases by constructing ``fringe''
solutions on \m348 itself; however, at the first epoch, \m348 was
relatively low in flux, so the local phase-reference calibrator
J0112+3522 was used instead.  Following the initial calibration, data
were self-calibrated and imaged in both AIPS and {\caps Difmap}
(Shepherd 1997), resulting in the final images discussed in this
paper.  Images made in AIPS used an intermediate weighting with a
Briggs robustness parameter of 0.  Gaussian fits were made to
component flux densities, sizes, and relative positions.  For the
extended lobes seen at 1.7~GHz, flux densities were determined by
integrating over the portions of the images where significant emission
was detected.

%%At the third epoch, cross-polarization data were obtained along with
%%the parallel hands of polarization.  Bandpasses were calibrated by
%%means of observation of the strong source J2253+1608 (3C~454.3), and
%%polarization leakage was calibrated using measurements of
%%cross-polarized fringes on strong sources.  The absolute polarization
%%position angle was determined by self-calibrating the source
%%J2136+0041 at 8.5, 15 and 22~GHz, then rotating all polarization
%%position angles by the amount required to correct this source to the
%%measured values given by {\bf (Jim??)}.

\subsection{VLBA Spectral Line Observations}

The high spatial resolution observations of the maser line were made
using all 10 antennas of the VLBA on June 10, 2000.  \m348\ was
observed for 9 hours over a 12 hour period.  Because the line was seen
in the March Effelsberg observations to be too broad to fit in a
single 16 MHz IF (full width at zero power $>$250 \kms), 2 IFs of 16
MHz each in both right and left circular polarizations were used,
overlapped somewhat in frequency.  256 channels were used in each IF,
resulting in a channel spacing of 125 kHz.  This corresponds to a
velocity resolution of 1.74 \kms.

Amplitude calibration was derived using the antenna gain curves and
measurements of system temperature as described in $\S$2.3.  The
calibrator source J0237+2848 (4C+28.07) was used as an initial fringe
finder, following which delay and rate solutions were calculated using
the bright, compact continuum source in \m348\ itself, using the AIPS
task FRING with a solution interval of 3 minutes.  The nearby source
J0136+4651 was observed every 40 minutes to ensure adequate bandpass
calibration.  This setup provided enough bandwidth to cover the line
and supply 20 line-free channels on the low frequency end of the line
which could be used to generate a clean ``map'' of \m348\ in {\caps
Difmap}.  This model of clean components was then used to remove any
amplitude offsets between the two IFs and between right and left
circular polarizations.  Unfortunately, no line-free channels were
available on the high frequency side of the line.  Following
calibration, the overlapping channels were removed and the 2 IFs were
joined together, using the AIPS task UVGLU, to yield a single cube of
174 channels covering 23 MHz.  Continuum subtraction was done in the
{\it u,v} plane using the AIPS task UVLSF.  Subsequent editing and
imaging of all data was done using {\caps Difmap}.  The line spectra
were then analyzed and fitted with Gaussian functions using the
Groningen Image Processing System, GIPSY (van der Hulst \etal\ 1992).

\section{Results}

\subsection{Effelsberg Monitoring of the \h2o\ emission}

\begin{figure}
\vspace{10cm} \includegraphics{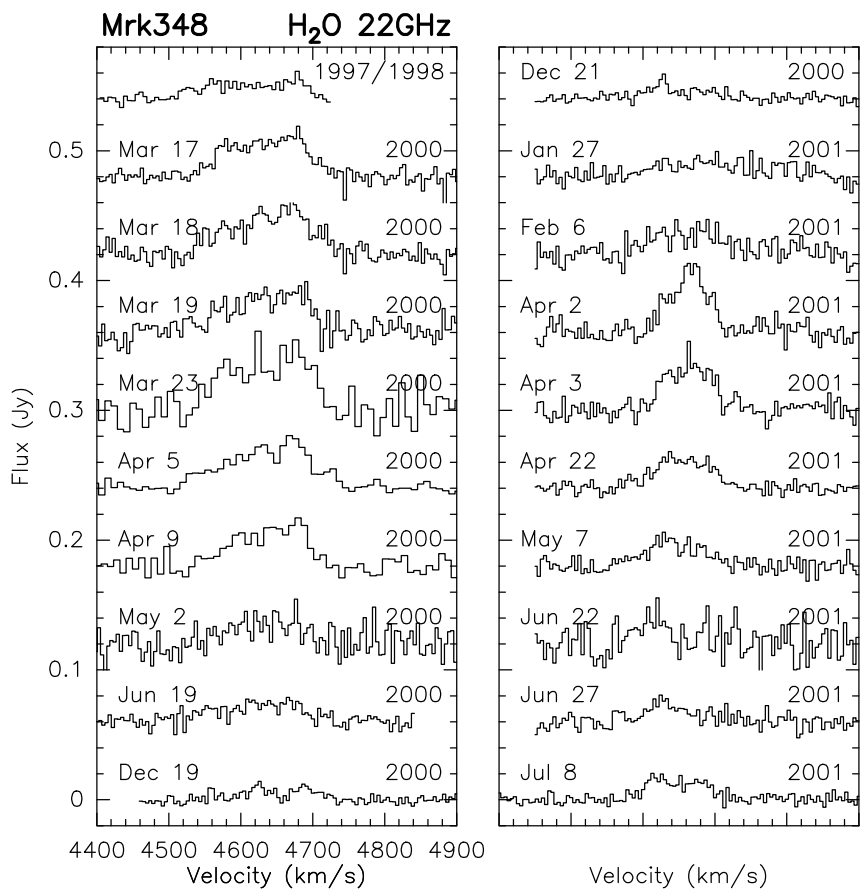} \figcaption{Single dish profiles
from the Effelsberg 100m telescope.  The peak flux in the line was
$\sim$40~mJy on April 9, but decreased to 9~mJy by June 19, 2000.  In
April 2001, the line peak again increased to 50~mJy, but the FWHM
decreased to $\sim$65 \kms.
\label{fig:100m}}
\end{figure}

The initial detection of the flaring maser in \m348\ using the
Effelsberg 100m telescope took place in 2000 March (shown in
Figure~\ref{fig:100m}, 2nd profile; see also Falcke \etal\ 2000).
Re-analysis of previous unpublished data on this source (shown in the
top profile, Fig.~\ref{fig:100m}) indicates that the maser was also
present but only marginally detectable in late 1997.  The \h2o\ maser
line in \m348\ is extremely broad, with a FWHM of $\sim$130 \kms,
though in many of the monitoring epochs the emission appears to
consist of 2 lines which can be tentatively fit with a broad Gaussian
function at \ab4609 \kms\ with FWHM\ab100 \kms\ and a narrower one at
\ab4678 \kms\ with FWHM\ab60 \kms.  The amplitudes of each component
vary significantly on very short timescales.  There are no detectable
high velocity components within 1500 \kms\ on either side of the
strong emission line (for the noise levels of the spectra, see
Fig.~\ref{fig:100m}; for upper limits over a wider velocity range, see
Falcke \etal\ 2000).

Monitoring of the \h2o\ emission through June 2000 showed that the
maser again decreased to its 1997/1998 level within 2 months, although
Xanthopoulos \& Richards (2001) report a value of 107~mJy on May 2
using MERLIN.  Our May 2 profile indicates a peak flux of
$\sim$16~mJy, consistent with our VLBA observations on June 10 (see
\S3.4).  Resumption of the monitoring program at Effelsberg in
December 2000 showed little change in the line flux, although the
component centered on \ab4609 \kms\ appears to be slightly stronger
than the higher velocity feature.  In February 2001, however, another
flare appears to have begun.  The observations made in early April
2001 (see Fig.~\ref{fig:100m}) show that the FWHM linewidth decreased
to 60\p3 km s$^{-1}$, and the velocity centroid increased from 4642 km
s$^{-1}$ to $\sim$4665\p1 km s$^{-1}$.  Thus it seems that only the
higher velocity line component was flaring.  Following this peak near
April 2, the line again began to decrease in flux.  Further
observations show that in June 2001, the low velocity component
underwent a smaller flare of unknown duration.  This behavior is
consistent with the scenario described in $\S$3.4 in which the higher
velocity component arises closer to the central engine of the radio
source.

\begin{figure}
\vspace{8cm}
\includegraphics{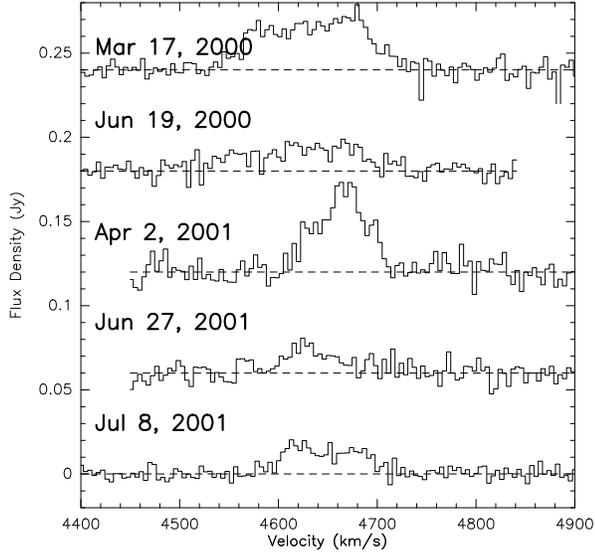}
\figcaption{5 \h2o\ spectra from the Effelsberg 100-m telescope chosen
  to illustrate the variations in the profiles seen on timescales of
  days of months, as discussed in text.
\label{fig:4profs}}
\end{figure}

Figure~\ref{fig:4profs} shows 5 spectra which were chosen to emphasize
the significant variations in the line profiles over our first 16
months of observations.  The first profile is the discovery spectrum,
where both components had a flux of around 30~mJy.  In the second
profile, taken within a week of our VLBA observations, both components
verged on undetectability and then, in the third profile, the higher
velocity component flared to twice the intensity of the year previous,
and the lower velocity component disappeared altogether.  This change
was reversed in June, shown in the fourth profile, when the lower
velocity component was present, though the intensity was low, and the
higher velocity component vanished.  Within two weeks, both components
were once again nearly equal.

\begin{figure}
\vspace{8cm} \includegraphics{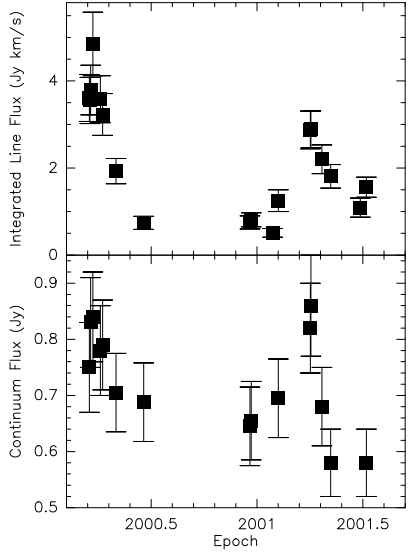} \figcaption{a) Integrated line flux
density vs. time; b) continuum flux density vs. time measured with the
Effelsberg 100-m telescope.  The line and continuum intensity can be
seen to vary concurrently.  The correlation between line and continuum
is $y=-5.36$\p$1.80 + 10.67$\p$2.46 \times x$, where y is the integrated line
flux density (Jy \kms), and x is the continuum flux density (Jy).
Thus a 1~mJy increase in continuum yields approximately 10~mJy \kms\
integrated line flux density.
\label{fig:linevtime}}
\end{figure}

Figure~\ref{fig:linevtime} shows the variation in line and continuum
flux density with respect to time.  Comparison shows that the
continuum flux varies loosely with the maser flux.  The implications
of this for the maser position with respect to the central engine are
discussed in $\S$4.2.

\subsection{Integrated Radio Continuum Flux Density Monitoring}

\begin{figure}
\vspace{10cm}
\includegraphics{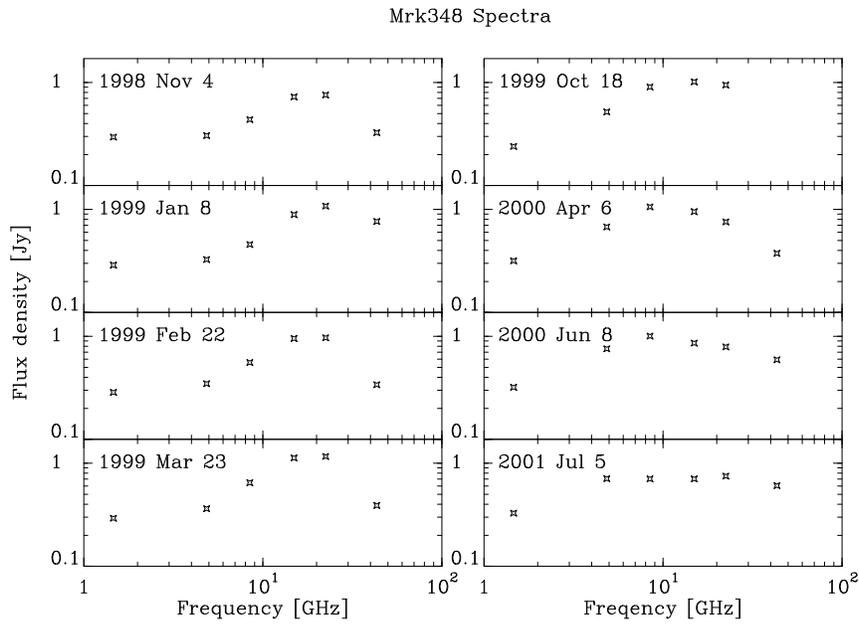}
\figcaption{Integrated flux density measurements made with the VLA at 6 frequencies.  The peak occurs at 22 GHz in the first 4 epochs, then gradually decreases in frequency over the next 20 months.
\label{fig:vlaspec}}
\end{figure}

Figure~\ref{fig:vlaspec} shows the continuum spectrum of \m348
measured at 8 epochs over 2.5 years using the VLA.  During the first
four epochs (1998 November to 1999 March), the spectrum of \m348 is
inverted at centimeter wavelengths and reaches its peak at
22~GHz. During the following epochs, this turnover dropped to lower
frequencies. At frequencies above the turnover, the spectrum gradually
flattens with time.  This behavior is similar to that in III~Zw~2,
where this type of evolution in turn-over frequency has been linked to
the expansion of synchrotron bubbles in a jet and decreasing
self-absorption.  This interpretation is further discussed in $\S$4.1.

\subsection{VLBA Continuum Imaging}

\begin{figure}
\vspace{8.5cm}
\includegraphics{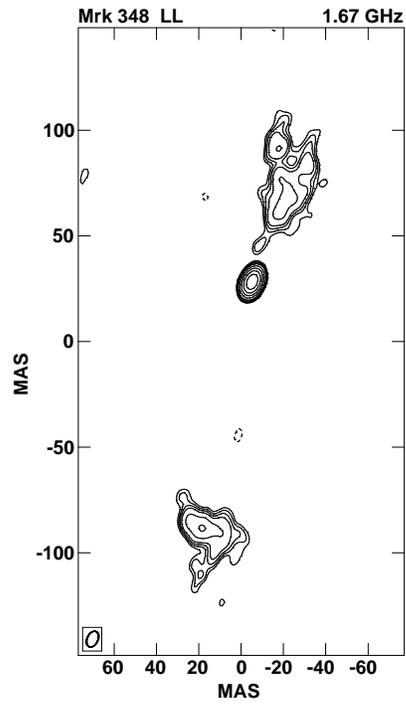}
\figcaption{1.6~GHz VLBA image of \m348, from 1997.10.  The x and y
  axes are offsets from the phase center in milliarcseconds.  Contours begin at
0.35~mJy~beam$^{-1}$, increasing by factors of $\sqrt{2}$ to
1~mJy~beam$^{-1}$ and by factors of 2 thereafter.  Negative contours
are shown dashed.  The peak flux density is 30~mJy~beam$^{-1}$.
\label{fig:18cm}}
\end{figure}

Figure~\ref{fig:18cm} shows the large-scale VLBI image of \m348 at
1.7~GHz, from 1997.10.  The core and outer lobes shown in
Figure~\ref{fig:18cm} were first imaged by Neff \& de Bruyn (1983);
the image shown here indicates that the northern and southern lobes
are quite diffuse and heavily resolved.  In fact, neither lobe is
detected at 5.0~GHz, with $4\sigma$ upper limits of
0.32~mJy~beam$^{-1}$.  This is consistent with the scenario in which
the emission in the lobes is smoothly distributed and has the spectrum
of optically thin synchrotron emission.  The absence of 5~GHz hotspots
may indicate that there is no powerful jet currently feeding the
lobes.  The total flux density detected in the northern lobe is
27~mJy, while that in the southern lobe is 17~mJy.  These values are
far below the respective 1.4~GHz values of 118~mJy and 63~mJy reported
by Neff \& de Bruyn (1983) who used shorter baselines in Europe,
indicating that the VLBA has resolved out much of the lobe structure.

\begin{figure}
\vspace{7cm} \includegraphics{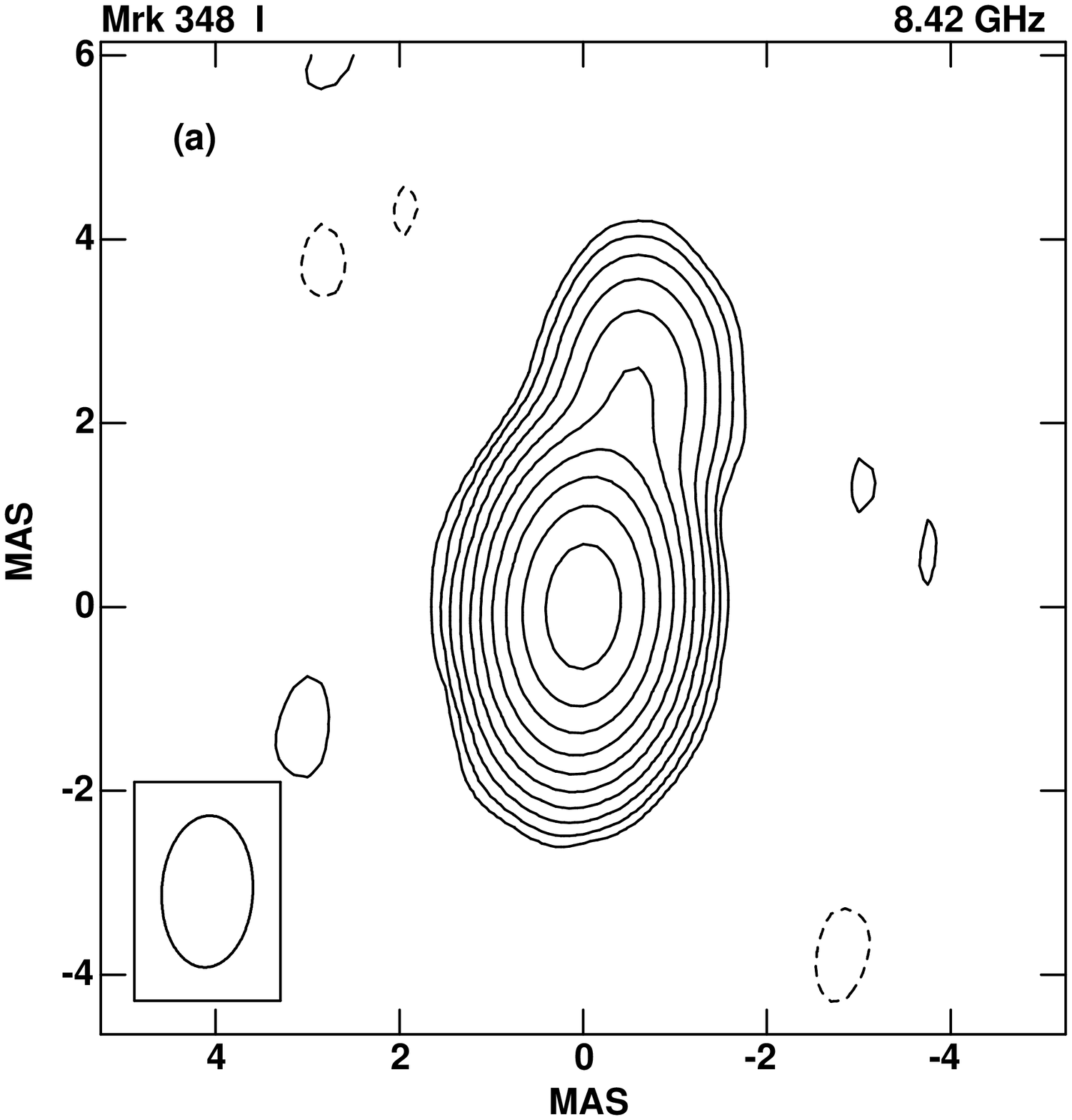} \includegraphics{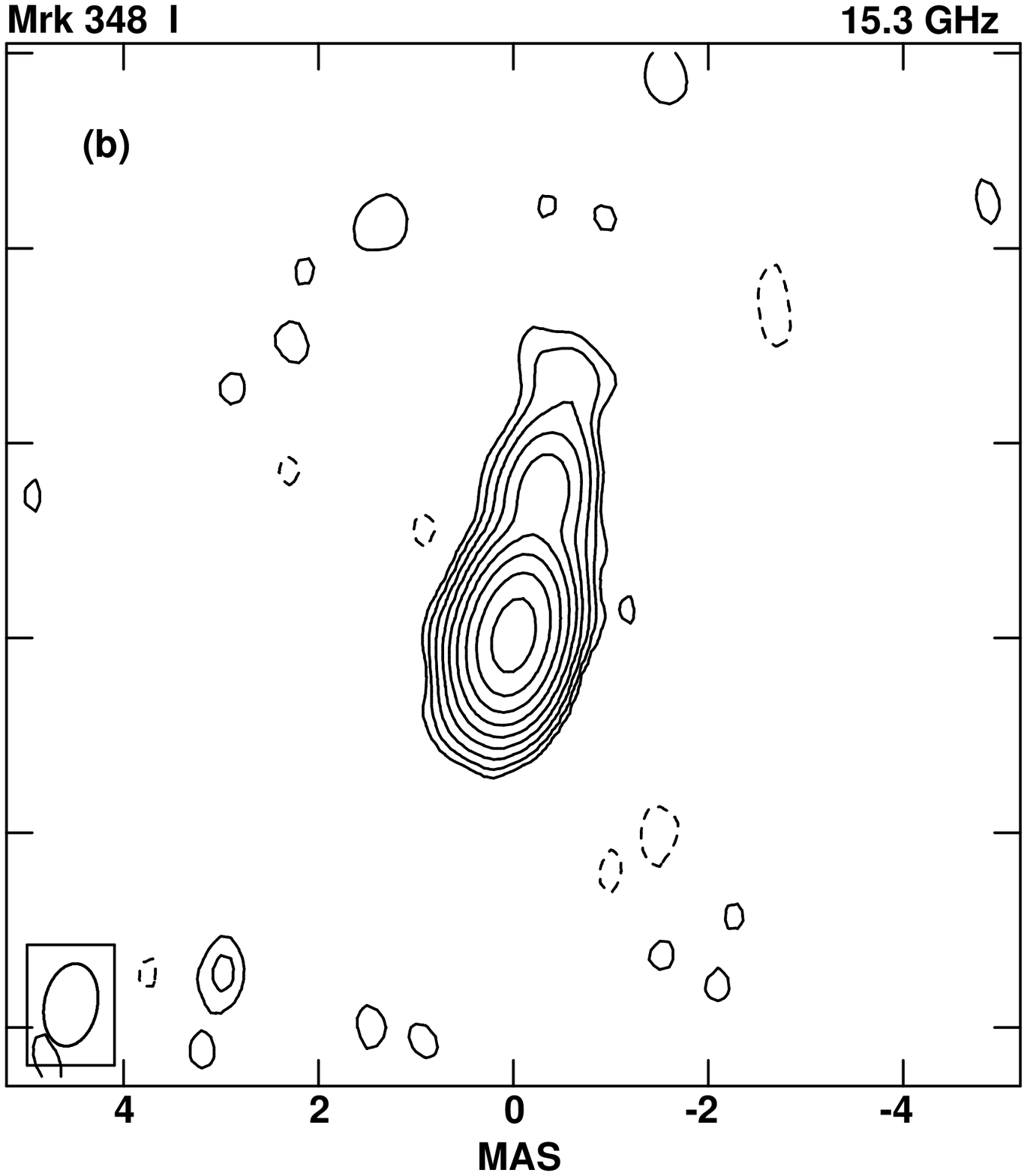} \includegraphics{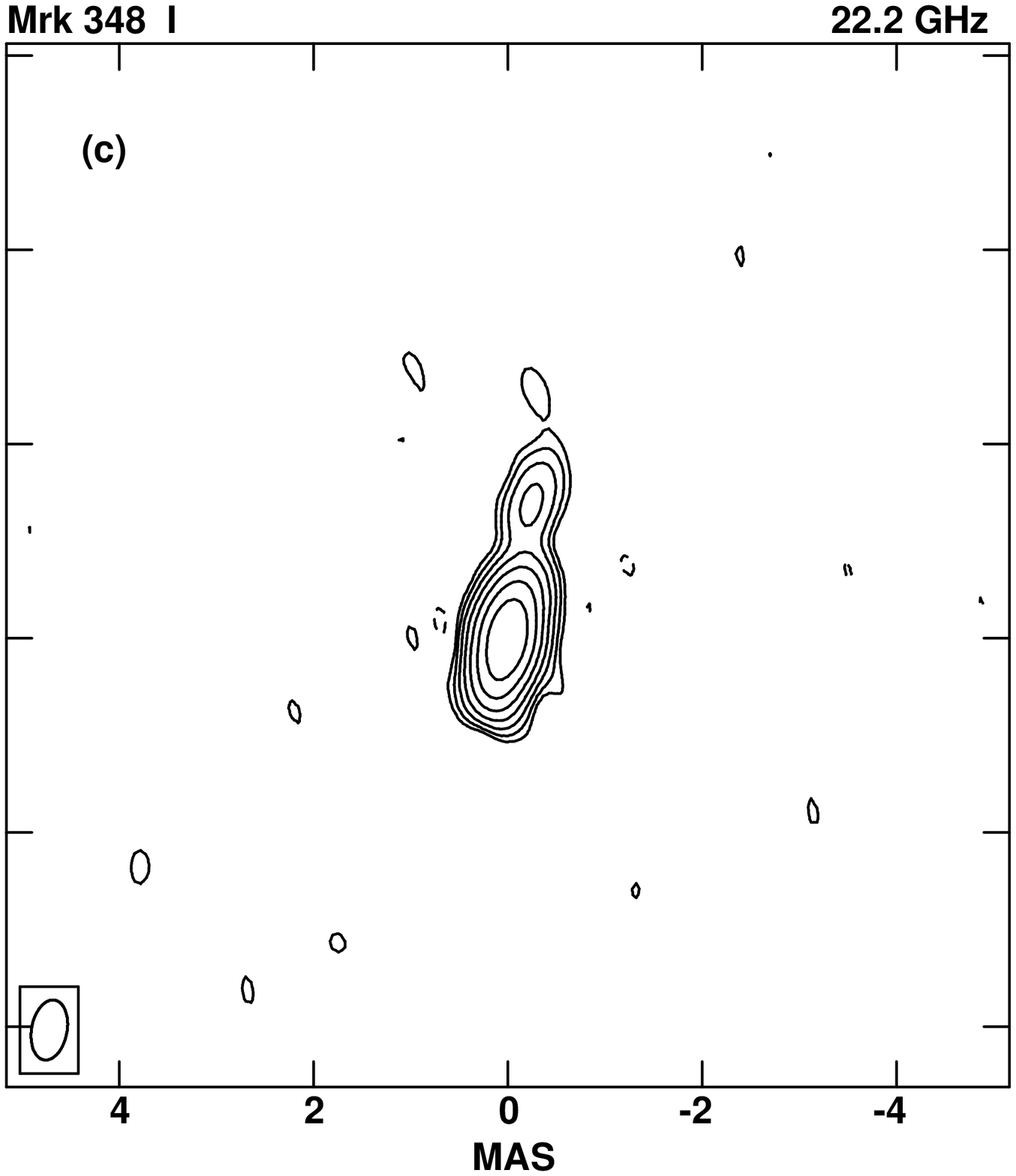} \figcaption{ Images
of \m348 from 2000.00, with contour intervals increasing by factors of
two from the lowest contour.  Beam sizes, given in
Table~\ref{tab:vlba-cont}, are shown in the lower left corners {\it
(a)}. Image at 8.42~GHz, with lowest contour at 1~mJy~beam$^{-1}$, and
a peak of 803~mJy~beam$^{-1}$.  {\it (b)}.  Image at 15.3~GHz, with
lowest contour at 2~mJy~beam$^{-1}$, and a peak of
764~mJy~beam$^{-1}$.  {\it (c)}.  Image at 22.2~GHz, with lowest
contour at 4~mJy~beam$^{-1}$ and a peak of 500~mJy~beam$^{-1}$.
\label{fig:VLBAcore}}
\end{figure}

\begin{figure}
\vspace{6.5cm}
\includegraphics{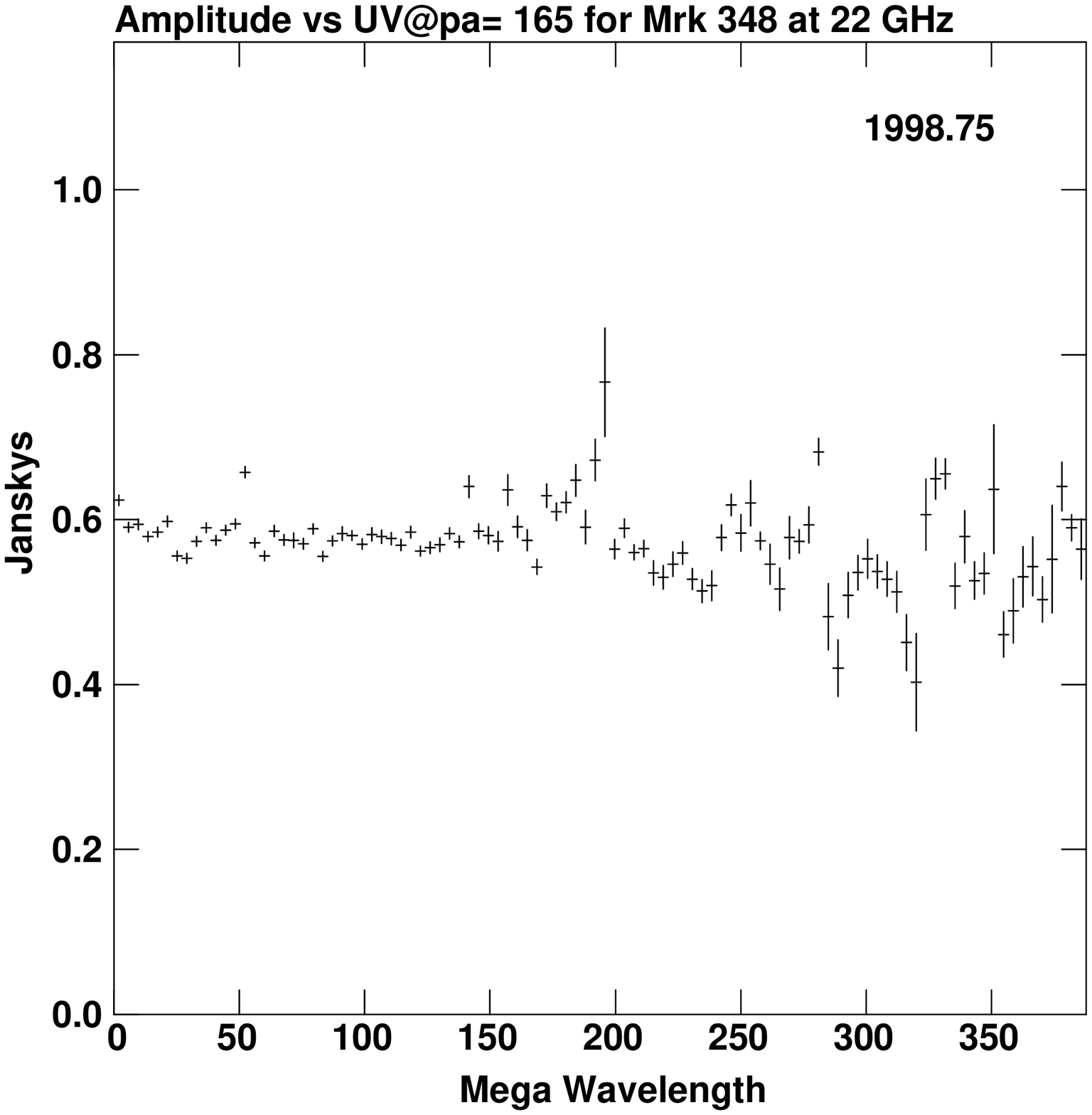}
\includegraphics{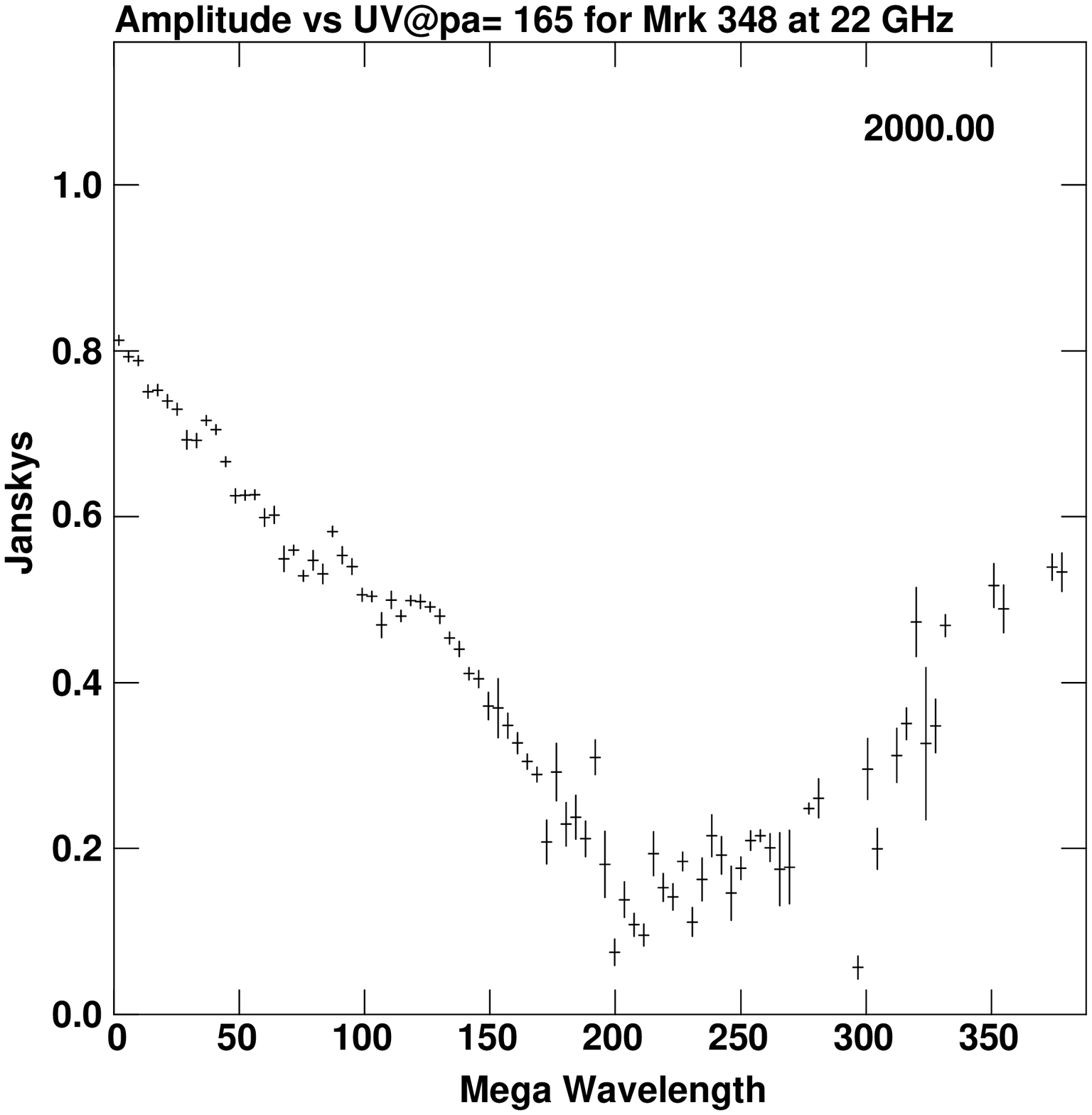}
\includegraphics{f7c.eps}
\figcaption{
22~GHz visibility plots along position angle $165^\circ$, from 
epochs 1998.75 (left panel), 2000.00 (central panel) and 2000.44
(right panel).
Note the clear minimum at a projected baseline
length of 210--240 million wavelengths in 2000.00, corresponding to a nearly
equal double source with a separation of 0.43--0.49~mas.  There is no
significant minimum in the visibilities at 1998.75.
\label{fig:visplot}}
\end{figure}

Figure~\ref{fig:VLBAcore} contains the VLBA images of the parsec-scale
emission from \m348 at 8, 15, and 22~GHz, from the data acquired at
epoch 2000.00.  In addition to the extended emission at $\sim 1.5$~mas
to the north-northwest of the core, first reported by U99, the core is
significantly resolved along the direction of the beam.  In fact, the
core appears to contain a new double source within the main peak.
This component is quite obvious in the visibility function;
Fig.~\ref{fig:visplot} shows the 22~GHz visibility amplitude and phase
plotted against projected baseline along position angle 165\deg\
(15\deg\ west of north), for data from 1998.75, 2000.00 and the
line-free channels from the maser observations on 2000.44.  This
figure shows that the dominant source (Component 1 in U99) has changed
from being barely resolved in 1998.75, with the flux decreasing
only slightly at longer baselines, to strongly resolved in 2000.44. In the
2000.00 data, there is a clear minimum in the visibility function,
which reaches almost to zero for a projected baseline separation of
210--240 million wavelengths.  This implies the presence of two
components with nearly equal intensity with a separation of
$\sim0.43-0.49$~mas. Multi-component Gaussian fitting in the image
plane confirms that the two strongest components are separated by 0.43
mas.  A somewhat larger separation is evident from the 22~GHz
continuum image made from the maser observations in 2000.44, where the
peak radio emission is now significantly displaced from the southern
edge of the core component.  The minimum in the visibility function is
at 160 million wavelengths, suggesting a separation between the two
components of $\sim0.64$~mas.  This is likely to be an overestimate
because the additional continuum component further to the northwest
causes the visibility minimum to move toward a shorter baseline length
(larger separation) than would be true for an isolated double source.
This yields an upper limit on the apparent separation speed of
$\beta_{\rm app}\le0.48$ between 2000.00 and 2000.44. The new component,
which was not separately detectable at 1998.75, seems to have appeared
concurrently with the flare in the nucleus of the galaxy. At 15~GHz,
Component~1 was reported by U99 to change from a 100~mJy unresolved
component at 1997.10 to a 550~mJy resolved component at 1998.75,
probably indicating the beginning of the emergence of the new
component.

\begin{figure}
\vspace{12cm} \includegraphics{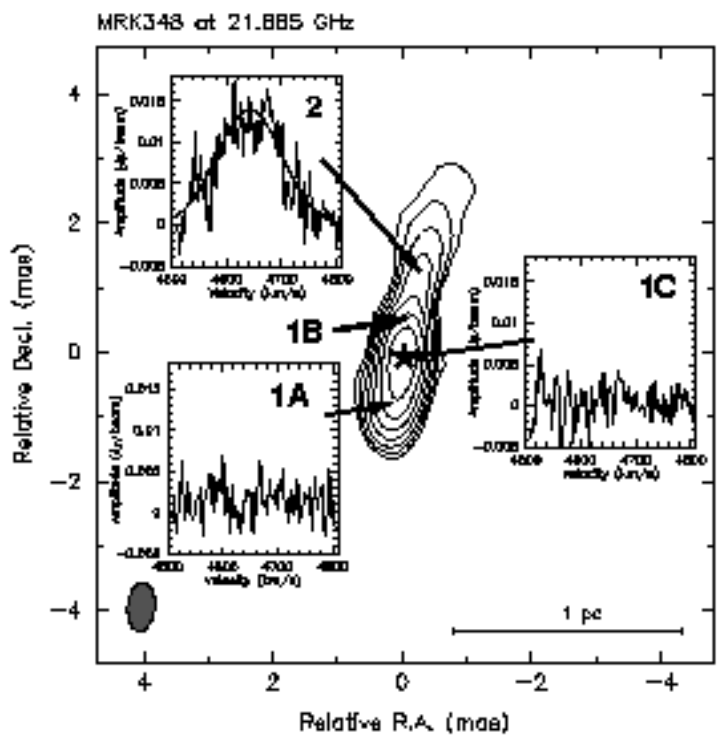} 
\figcaption{\h2o line profiles
toward \m348, superimposed on the continuum map made from 20 line-free
channels extracted from the low frequency end of the observed frequency range.
The data were taken in 2000.44.  The RMS noise in the line profiles is
\ab4~mJy~beam$^{-1}$~channel$^{-1}$.  The continuum image is naturally
weighted and the lowest contour is 5 ~mJy. The RMS noise in the
continuum map is less than 1~mJy~beam$^{-1}$.  The suspected position
of the core is indicated with an asterisk, and the other components
discussed in the text are labelled accordingly.
\label{fig:profplot}}
\end{figure}

Table~\ref{tab:vlba-res} contains the results of Gaussian fits to the
core components at various frequencies and different epochs.  Within
the central few milliarcseconds of the source, components are numbered
from South to North; because of the newly emerging components,
Component 1 of U99 has been split into first two components, 1A and
1B, for the 2000.00 data, and then a third component designated 1C
that appears between them in the 2000.44 image made by averaging the
line-free channels in the spectral-line data.  This new component may
indicate the re-emergence of the ``core'' component previously
referred to as Component 1 as 1A and 1B have separated sufficiently to
``expose'' 1C. Its location is marked in Figure~\ref{fig:profplot}.

Adding the third epoch at 15~GHz, measuring all components relative to
the position of 1A, the southernmost component, we find that the
proper motion of component 2 between the first and third epochs is
$0.075\pm 0.004$ ~mas~yr$^{-1}$ ($1\sigma$ error quoted), consistent
with U99 who found a proper motion of $0.075\pm 0.035$~mas~yr$^{-1}$
between the first two epochs.  This corresponds to $\beta_{\rm app} =
0.076\pm 0.004$.  The relative speeds of components 1 (or 1A) and 2 at
8.4 and 22~GHz have not been assessed, because component 1 at 8.4~GHz
is affected significantly by apparent synchrotron self-absorption,
while component 2 is well-resolved at 22~GHz at 2000.0, and has no
isolated component peak.  We note that component 2 has brightened
and component 3 has appeared at 15 GHz between 1998.75 and 2000.00,
implying that they may be responding to the flare in the nucleus. 

\begin{figure}
\vspace{8cm}
\includegraphics{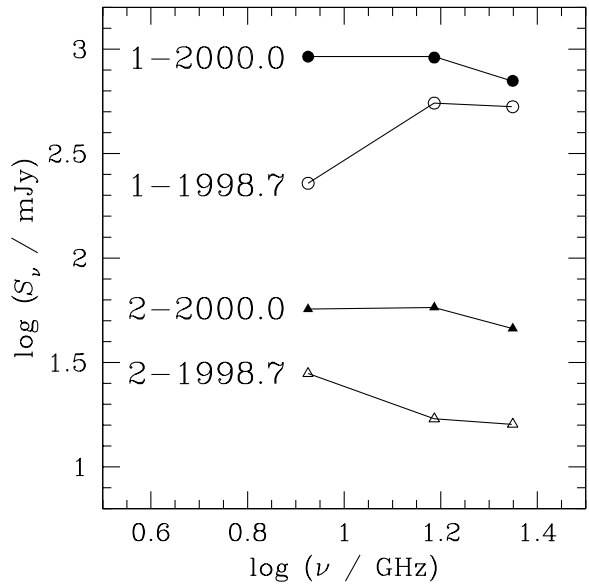}
\figcaption{
Spectra of components 1 and 2 of Mrk 348,
at epochs 1998.75 and 2000.00.  All sub-components of
1 have been summed in this plot.  Error bars of
$\pm$ 10\% (including fitting errors) are slightly
larger than the symbol sizes, and are not shown in order
to reduce clutter.
\label{fig:spectra}}
\end{figure}

Figure~\ref{fig:spectra} shows the radio spectra of components 1 (all
sub-components added together) and 2 at 1998.75 and 2000.00.  The
radio flare at 8.4~GHz appears to lag the flares at 15 and 22~GHz
(also shown in the integrated flux measurements in
Fig.~\ref{fig:vlaspec}), perhaps due to optical depth effects; in this
regard, at 2000.00, component 1A is $\sim 0.12$~mas closer to 1B at
8.4~GHz than at the higher frequencies.  The spectrum of component 1
seems to have a maximum near 15~GHz at both epochs, and its
sub-components also peak near 15~GHz. Given the sub-parsec sizes of
the components, synchrotron self-absorption is likely.  However, since
all components show a peak relatively near the same frequency at
different epochs, the frequency of the peak emission of component 1
was probably between 5 and 15~GHz at 1997.10.  This may imply that the
radio emitters are free-free absorbed by a common parcel of foreground
gas, perhaps in the circumnuclear torus that was discussed by U99.

\subsection{VLBA Spectral Line Imaging}

Figure~\ref{fig:profplot} shows three line profiles toward \m348 taken
in 2000.44.  The maser emission is clearly seen to lie along the line
of sight to component 2, the fainter continuum component, rather than
components 1A, 1C and 1B which comprise the brightest region of the
continuum source.  The Gaussian fit to the line shown in the upper
left profile has an amplitude of 14\p2~mJy and an integrated flux of
2.11\p0.34 Jy/beam/\kms, indicating that all of the flux measured in
the Effelsberg May 2 observation (2.12\p0.18 Jy/beam/\kms\ with a peak
of 16 mJy) has been recovered.  The FWHM is 139\p11 \kms\ centered on
4641\p2.2 \kms, consistent with the single-dish measurements and
redshifted by 134 \kms\ with respect to the systemic velocity.  A
tentative 2 component fit to the data yields a narrower line at
4682\p3 with FWHM\ab60 \kms\ and amplitude 9\p1~mJy, and a broader
line at 4617\p4 with FWHM\ab100 \kms\ and amplitude 11\p1~mJy, again
consistent with our single-dish measurements.

\begin{figure}
\vspace{8cm} \includegraphics{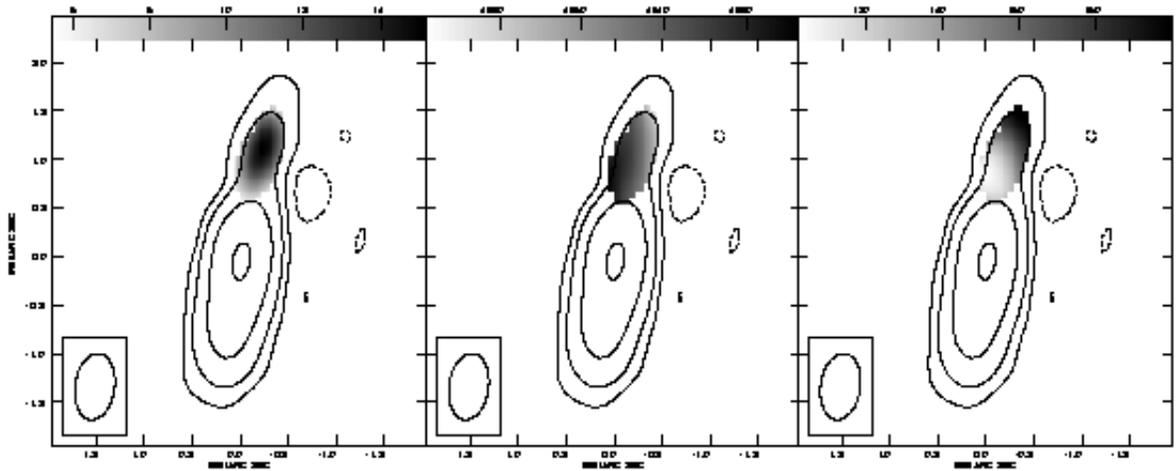} \figcaption{Amplitude (left, in mJy),
velocity (center, in \kms) and FWHM (right, in \kms) fields of the
maser emission obtained by fitting a Gaussian function at each pixel
are shown in gray scale, superposed on the 22 GHz radio continuum
image displayed as a contour plot.  Although the maser emission is
unresolved, the slight velocity gradient seen in the central panel, in
conjunction with the gradient in FWHM, can be interpreted to mean the
narrower, higher velocity line component arises closer to the central
engine.
\label{fig:pixbypix}}
\end{figure}

The plots in Figure~\ref{fig:pixbypix} show Gaussian fits to the data
at each pixel.  Pixels where the signal to noise ratio in the line was
less than three have been blanked.  Maser emission is seen only toward
the northern jet.  This emission is unresolved at our angular
resolution of 0.42$\times$0.76 mas, corresponding to a linear size of
less than 0.25 pc.  The central panel in Figure~\ref{fig:pixbypix}
shows a velocity gradient that, in conjunction with the gradient in
FWHM depicted in the right panel, appears to show that the narrower,
higher velocity emission component described above arises to the
southeast of the slightly broader, lower velocity line.  While this
gradient by itself should not be over-interpreted as the masing region
is unresolved, the possibility that the higher velocity component may
be caused by masing in gas closer to the central engine of the source
is supported by the fact that the higher velocity component was seen
to flare in April 2001, 80 days before a similar flare in the lower
velocity component was observed, as described in $\S$3.1.  Although we
do not have high enough time resolution in our monitoring program to
ascertain exactly when either flare began, we can still estimate the
maximum distance between the two regions responsible for these masers,
based on the distance covered in 80 days at light-speed, to be
2.07$\times$10$^{17}$ cm, or 0.07 pc.  This distance is well within
the size restrictions imposed by the maximum linear size of the maser
source.  Much higher time resolution is needed to measure an accurate
time delay.

\section{Discussion}

Megamaser sources in which the mechanism causing population inversion
is thought to be an interaction between the radio jet and a molecular
cloud can provide detailed information about the conditions in the
central parsecs of active galactic nuclei (AGN).  The study of these
``jet masers'' can yield information about the molecular clouds in the
interstellar medium (ISM) of the host galaxy, because population
inversion in water molecules following a shock in a molecular cloud
implies a fairly narrow range of temperatures and pressures in the
pre-shock gas ({\it e.~g.} Elitzur 1995).  These sources can also
yield important information about the evolution of jets and their
hotspots.  Systematic reverberation mapping may allow us to determine
jet speeds, variations in source sizes and positions, actual distances
between the various targets, an estimate of inclination by comparing
these distances with projected angular separations, and the relation
of the maser components to continuum flares. Although we have only a
small number of epochs so far, we can address the evolution of the
continuum morphology on mas scales, analyze the physical properties of
the gas giving rise to maser emission, and hypothesize about future
prospects of jet maser detections.

%%If the core of the radio source is responsible for an
%%increase in continuum flux density, monitoring of the flux densities
%%of the continuum and the line emission can provide estimates, through
%%reverberation mapping, of the speed of the material in the jet,
%%particularly in sources like \m348, where the jet appears to lie close
%%to the plane of the sky.  If, on the other hand, an increase in
%%continuum flux density is caused by the brightening of the hotspot or
%%working surface in the jet as it impacts a denser part of the ISM,
%%then the onset of a continuum flare and a maser flare should be nearly
%%simultaneous.  High angular resolution studies of these sources allow
%%us to put constraints on the size of the excited regions in the
%%molecular clouds, and to determine the distance of the maser emission
%%from the central engine.  We can determine whether the maser emission
%%is coincident with the termination of the jet and what its relation is
%%to the region causing the continuum flare.  This in turn can tell us
%%if studies of the proper motions of the jet components are indeed
%%indicators of the speed of the material in slow jets, or whether they
%%simply show the rate at which the shock caused by the impact of the
%%jet material propagates through a dense cloud in the ISM.

\subsection{Continuum Evolution on mas Scales}

VLBA observations of \m348 have thus far shown only sub-relativistic
expansion (U99).  The initially GHz-peaked integrated spectrum of
\m348 (Fig.~\ref{fig:vlaspec}) can be interpreted as an enhancement of
one spatial region in a conical jet where the peak frequency is
inversely related to the spatial scale (see Blandford \& K\"onigl
1979; Falcke \& Biermann 1995). The enhancement could be caused by the
formation of compact hotspots as a consequence of the jet hitting a
dense target, such as a molecular cloud or torus, or the appearance of
new bright components emanating from the central engine. The fact that
the turnover frequency of the spectrum was dropping at the end of 1999
can be interpreted as expansion of the source and supports the
interpretation of synchrotron self-absorption for the inverted
spectrum. The flattening of the spectrum at higher frequencies
suggests that scales even more compact and therefore probably closer
to the nucleus brighten.  This behavior can be compared to the
spectral evolution in the recent outburst of the Seyfert 1 galaxy
III~Zw~2. III~Zw~2 showed a self-absorbed synchrotron spectrum with a
turnover frequency of 43~GHz (Falcke \etal\ 1999). The turnover
frequency stayed constant during the outburst, and the source showed
no expansion on sub-pc scales. Then the spectral peak dropped quickly
to 15~GHz within a few months. This phase of rapid spectral evolution
in III~Zw~2 was accompanied by a superluminal expansion of the source
on sub-pc scales (Brunthaler \etal\ 2000).

The evidence indicates that the center of activity in \m348\ is likely
to be located between components 1A and 1B.  At an epoch prior to
1998.75, jet components were ejected simultaneously in the approaching
and receding jets, leading to both the continuum flare and the new
flaring activity of the maser.  We base this interpretation both on
the fact that the maser emission is redshifted with respect to the
systemic velocity of the source, indicating that it most likely arises
in the receding jet (see $\S$ 4.3), and on the fact that
%%, while the extended jet itself is not
%%visible on either side of the source (meaning that beaming should not
%%play as important a role in determining the locations of the hotspots
%%as collisions with material in the host galaxy), 
we expect to
see a detectable corresponding approaching jet component in any radio
source where we detect a receding jet component. Thus it seems
unlikely that the central engine could be located at the southern
extremity of the radio source as had been inferred by U99.
Figure~\ref{fig:profplot} shows the image made from 20 line-free
channels at the low frequency end of our observing band in the 2000.44
observations.  Our inferred position of the core is marked with an
asterisk.

Limits can be placed on the separation speed of the new components, 1A
and 1B.  A firm lower limit is derived by assuming that at least one
of the new components did not exist at 1997.10, before the radio
flare.  Therefore, the relative separation of 0.43~mas in 2000.00
(\S3.3) occurred in less than 2.90~yr, or at a rate of at least
0.15~mas~yr$^{-1}$, twice the apparent speed of component 2 during the
same period.  An improved lower limit can be derived from the single
Gaussian source size of 0.16~mas for Component 1 at 1998.75.  This
indicates that the separation of 1A and 1B was likely to be no more
than 0.16~mas at that epoch, consistent with the lack of a visibility
minimum out to nearly 400 million wavelengths (Fig.~\ref{fig:visplot})
giving a lower limit of 0.22~mas~yr$^{-1}$ between 1998.75 and
2000.00.  The resolution of component 1 at 1998.75 indicates that 1A
already existed, so the upper limit to the apparent speed is 0.43~mas
in 1.25~yr, or 0.34~mas~yr$^{-1}$.  Converting angular motions to
linear speeds, we find $0.22 \lesssim \beta_{\rm app} \lesssim 0.34$
for the apparent relative speeds of components 1A and 1B between
1998.75 and 2000.00.  This is also consistent with the upper limit of
$\beta_{\rm app}\le0.48$ between 2000.00 and 2000.44.  Although
component 2 appears to have sped up between 2000.00 and 2000.44,
perhaps having emerged from the far side of the molecular cloud, the
apparent higher speed of 1B relative to 1A implies that it would have
caught up with component 2 at approximately 2001.3.  Further VLBA
continuum monitoring experiments have been undertaken to determine
whether this has occurred.

\subsection{Molecular Gas in the Central Few Parsecs}

One concern raised by the scenario of a molecular cloud within a
projected distance on the order of one parsec from a supermassive
object is whether the gas density is high enough to keep the cloud
from being disrupted by tidal forces.  Using 2$\times$10$^8$ \solmass\
as the central mass of Mrk~348 (Nishiura \& Taniguchi 1998), and an
approximate distance of one parsec, we can calculate whether this
cloud is likely to remain intact.  If we assume that the gas is
roughly virialized, but infalling so that the kinetic energy is
slightly less than the gravitational potential energy, then the upper
limit on the circular velocity ($v_{\rm circ}$) around a central
massive object of 2$\times$10$^8$ \solmass\ at a distance of 1 pc is
\ab900 \kms.  Using this upper limit, we can then estimate the average
density ($\rho_{\rm cloud}$) following the reasoning of Stark \etal\
(1989).  They estimate the density needed to keep individual parcels
of gas bound to a central cloud core near a large central potential to
be

$<\rho_{\rm cloud}>$ $\approx$ 3/2$\pi$G $\times$ $v_{\rm
circ}^2$/R$_0^2$ cm$^{-3}$

\noindent where R$_0$ is the distance of the cloud from the dynamical
center.  Using this relationship, we obtain $<\rho_{\rm
cloud}>\approx$4$\times$10$^9$ cm$^{-3}$.  This high value is regarded
as an upper limit for the required density of the pre-shocked gas in
the masing region for two reasons: first, the trajectory of the
molecular cloud is not known but a stable virial orbit in a cloud this
close to the central engine seems unlikely; and second, the masing
region is known to be $<$0.25 pc in extent, and so could correspond to
a less dense region or regions in a larger molecular cloud.  This
upper limit is thus reasonably consistent with the highest pre-shock
density one would expect before maser quenching sets in, around
2$\times$10$^8$ cm$^{-3}$ for high velocity shocks ({\it e.~g.}
Elitzur 1995).  This is also within the range of gas densities
required to provide enough shielding to keep the gas in a molecular
phase, as predicted by theoretical models based on NGC~4258 (Maloney
2002; Maloney, Hollenbach \& Tielens 1996).  Within 1 pc of an x-ray
source yielding 10$^{43}$ ergs s$^{-1}$, a depth of only \ab10$^{15}$
cm ($<$10$^{-3}$ pc) of shielding gas having a density of
10$^7-$10$^8$ cm$^{-3}$ is needed, well within the size of the masing
cloud estimated by the time lag between flaring components in \S3.4.

\subsection{Physical Conditions in the Masing Region}
\subsubsection{Shock Velocities and Gas Densities}
\begin{figure}
\vspace{10cm}
\includegraphics{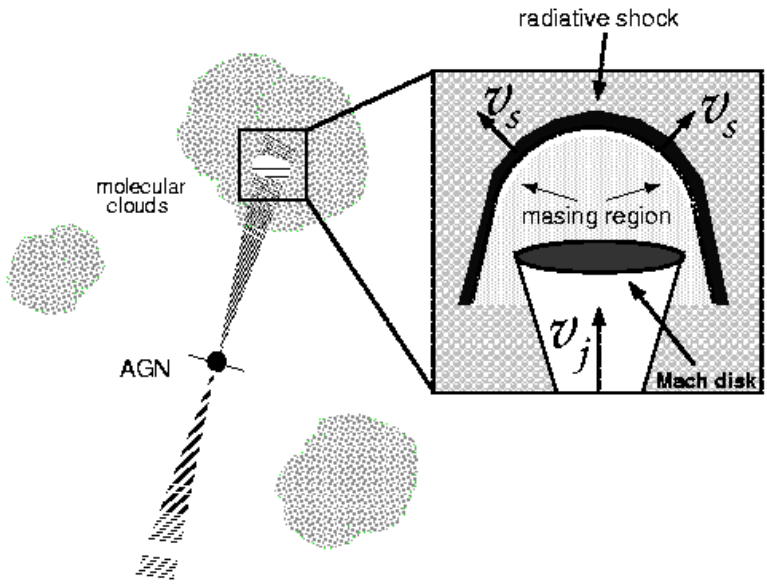}
\figcaption{A cartoon model of the expanding bubble caused by the jet
material impacting the molecular cloud.  The maser emission arises
within the region surrounded by the radiative shock.
\label{fig:model}}
\end{figure}

The jet in \m348\ is probably relativistic but the shock velocity
which gives rise to the maser emission is determined not only by the
jet velocity, but also by the relative densities of the jet and
ambient material.  The shock front is at the outer edge of an
expanding bubble or cocoon driven into the cloud by the jet.  A
cartoon model of this scenario is shown in Figure~\ref{fig:model}.
The masing region is located within or immediately behind the
radiative shock.  This situation is analogous to that found in high
mass star forming regions, where collimated outflow from a young
stellar object drives an overpressured cocoon into a dense ISM (see
{\it e.~g.} Reid \etal\ 1995; Mac Low \etal\ 1994).  The morphology of
such a bubble in extragalactic sources was explored by Scheuer (1974),
who determined that the bulk of the expansion would take place in the
direction of jet propagation, but that the bubble would also expand
laterally at slower speeds.  The velocity of the shock front in the
direction of propagation of the jet ($v_{\rm top}$) is related to the
jet velocity ($v_{\rm j}$) and the ratio of the densities of the jet
($\rho_{\rm j}$) and the ambient material ($\rho_{\rm 0}$) by $v_{\rm
top}=v_{\rm j}$($\rho_{\rm j}$/$\rho_{\rm 0}$)$^{1/2}$ ({\it e.~g.}
Elitzur 1995).

In \m348, the jet axis seems to be fairly close to the plane of the
sky, as indicated by the optical observations of Simpson \etal\ (1996)
and the flux densities of outer jet components (Neff \& de Bruyn 1983)
and also by the fact that the axis of the radio jet might be close to
the plane of the galaxy, which has an inclination angle of \ab16\deg\
(Braatz, Wilson \& Henkel 1997), in order for the jet to be hitting a dense
molecular cloud.  Further evidence for the small angle between the
radio axis and the plane of the galaxy has recently been presented in
the thorough study by Ant\'on \etal\ (2002).  The orientation of the
jet close to the plane of the sky should result in shocks with the
right orientation to yield strong masers along our line of sight.
This situation is analogous to the Galactic masers in star-forming
regions, where lower Doppler velocities are correlated with stronger
masers (Elitzur, Hollenbach \& McKee 1992).  Because all of the
detected maser emission is redshifted with respect to the systemic
velocity, we assume that the jet component toward which the maser
emission is seen is probably the receding jet, and the redshift is
caused by the entrainment of gas in this jet which is directed somewhere
between the plane of the sky and along the line of sight away from us.
In order to maintain velocity coherence, we suspect that the shock
velocity in the direction of jet propagation (the ``head'' of the
bubble) should be at least equal to the line of sight component of
entrainment velocity.  If the jet is close to the plane of the sky,
the total shock velocity might be much higher.  The maximum shock
velocity expected is \ab300 \kms, since anything higher than that will
destroy the dust grains which are required to provide condensation
nuclei in the post-shock gas for the hydrogen to coalesce into H$_2$
and then to form \h2o\ (Elitzur, Hollenbach \& McKee 1989).  If we
assume that the maser emission comes from the head of the expanding
bubble and $v_{\rm j}$ is between 0.1c and 0.5c (a range covering the
measured expansion speed of the jet components), then a density
contrast of 10$^6$ between the jet and the ambient material is
required.  For $\rho_{\rm 0}$\ab10$^8$ cm$^{-3}$ based on the
requirements described above to preserve the cloud (\S4.2), this
implies $\rho_{\rm j}$\ab100 cm$^{-3}$.  For a higher density jet, the
dust would be disrupted in the region immediately behind the shock at
the head of the bubble, and so the maser emission would have to arise
from regions where the shock front is expanding at slower speeds
laterally into the cloud.  If the two velocity components seen in the
line profiles are representative of masing arising from the shocks on
``front'' and ``back'' sides of the bubble from our viewpoint, then a
lower limit on the shock velocity can be estimated based on the line
of sight components of the separation velocity.  The centroids of the
main line components are separated by \ab 70 \kms\, yielding a lower
limit on the shock velocity of \ab35 \kms.  This lower limit is still
consistent with the range of velocities needed to produce population
inversion in gas with a pre-shock density between 10$^6$ and a few
times 10$^8$ cm$^{-3}$.

\subsubsection{Chemical Considerations}

Although few other molecules have been searched for since the onset of
the continuum flare in \m348, we have made a deep search for ammonia
with the Effelsberg 100m telescope.  We do not detect any ammonia
(NH$_3$) in absorption toward the continuum source, with 1$\sigma$
upper limits of between 17 and 10~mJy for the (1,1) through (4,4)
lines with 1 \kms\ channel spacing (Falcke \etal\ 2000).  This is
perhaps not surprising in view of the conditions in the post-shock
gas.

H$_2$O and NH$_3$ are hydrogenated molecules. These are, as a
consequence of the high mobility of atomic hydrogen on cold surfaces,
abundant on dust grain mantles (e.g. van Dishoeck \& Blake 1998). A
moderate velocity C-shock (velocities ranging from 0 to 60 \kms), as
well as a more violent J-shock (velocities $\ge$60 \kms)
that disrupts icy dust grain mantles without destroying the dust
entirely, can cause a significant increase in the gas phase abundances
of H$_2$O and NH$_3$.  Evaporation temperatures are $\sim$100\,K for
both molecules.

H$_2$O abundances appear to be [H$_2$O]/[H$_2$] $\sim$
10$^{-9}$--10$^{-7}$ in cool molecular cloud cores (e.g. Melnick
\etal\ 2000; Snell \etal\ 2000a,b,c; Neufeld \etal\ 2000b), 10$^{-6}$
in stellar outflow sources (e.g. Neufeld \etal\ 2000a), and 10$^{-4}$
in warm shock heated gas (Melnick \etal\ 2000). At temperatures of
several hundred to a few thousand degrees, common in C-shocks or in
the radiative region behind J-shocks, reactions with energy barriers
become efficient and direct gas phase formation of H$_2$O takes place
via O + H$_2$ $\rightarrow$ OH + H and OH + H$_2$ $\rightarrow$ H$_2$O
+ H.

In the case of NH$_3$, quiescent dark clouds have typically
$N$[NH$_3$]/$N$[H$_2$] $\sim$ 10$^{-7}$ (e.g. Benson \& Myers 1983;
their Table~5), but abundances become 10$^{-6}$--10$^{-5}$ in hot
cores (e.g. Mauersberger, Henkel \& Wilson 1987; Hermsen, Wilson \& Bieging 1988; Cesaroni
\etal\ 1994), where dust grain evaporation is believed to modify the
chemical constituents of the gas. Unlike H$_2$O, gas phase NH$_3$
formation appears not to be drastically accelerated in the warm post
shock gas (e.g. Ho \& Townes 1983).

H$_2$O is less rapidly destroyed by photodissociation than NH$_3$
(e.g.  Crovisier 2001) and has a higher energy threshold for
photodissociation ($\sim$6.6 versus 4.1~eV; Crovisier 1989; Suto \&
Lee 1983). Thus NH$_3$ should be destroyed more rapidly by the
ultraviolet radiation that is expected to arise during the hot
adiabatic phase of a J-shock. We thus find two qualitative scenarios:
Behind a C-shock, where the bulk of the gas remains molecular and
where no additional UV-radiation is produced, H$_2$O {\it and} NH$_3$
abundances should be enhanced due to the release of grain mantle
material into the interstellar medium. In more violent J-shocks,
however, where gas phase molecules get destroyed, H$_2$O re-formation
proceeds much faster, since potential barriers do not inhibit its
direct gas-phase formation and since photodissociation by UV radiation
from the shock affects predominantly NH$_3$. Thus the non-detection of
NH$_3$ implies a J-shock scenario, consistent with the argument above
that relativistic jets from the nuclei of active galaxies should have
the potential to create fast J-shocks in molecular clouds.
%%We thus qualitatively conclude that quasithermal NH$_3$ line radiation 
%%(potentially visible in absorption against the radio continuum of the jet) 
%%should not be detectable.

\subsection{Prospects for Future Maser Detections}
Although many water maser surveys have been conducted on Seyfert
galaxies with low detection rates (Braatz \etal\ 1997 and refs. therein),
the newly discovered ``jet masers'' suggest a new criterion for
selecting targets for much deeper searches.  In sources where the
orientation of the radio jets with respect to the disks of the spiral
galaxies is known, a small separation angle makes them worthy of a
much longer integration than has been afforded individual sources in
previous large samples.  New receiver technology also makes it
possible to detect much weaker masers than in the past.  In
particular, the masers in NGC 1052 and \m348\ would not have been
detected but for exceptionally sensitive observations with a flat
baseline (as are now routinely available at 22~GHz at the Effelsberg
100m telescope).  We are currently undertaking a survey of target
sources which have been selected from a collection of Seyfert galaxies
in which both the inclination of the host galaxy and the linear extent
of the radio source are known (Nagar \& Wilson 1999).  To maximize our
detection rate of jet masers we have chosen nearby active galaxies
for which there is a high probability of strong interaction between
radio jets and galaxy ISM. These sources are Seyfert galaxies with a
face-on ($i<$35\deg) galaxy disk in optical observations and extended
radio structures in VLA or VLBA observations, indicating that both the
disk of the galaxy and the radio jet should be fairly close to the
plane of the sky.  This combination of geometries increases the
probability that the radio jet lies close to the disk of the
galaxy. Indeed, three of the four known jet maser sources,
NGC~1068, NGC~1052 and \m348\ were among the galaxies selected from
the parent sample using the above criteria, suggesting that this
method of selection should yield a much higher detection rate than has
been achieved in past surveys.

\section{Conclusions}

During early 2000, the H$_2$O emission toward \m348\ showed a dramatic
intensity increase which followed a significant increase in the flux
of the nuclear radio continuum source in late 1998.  The unusual line
profile led us to suspect that this source, like NGC~1052 (Claussen
\etal\ 1998), might belong to a class of megamaser galaxies in which
the amplified emission is the result of an interaction between the
radio jet and an encroaching molecular cloud, rather than occurring in
a circumnuclear disk (Falcke \etal\ 2000).  Analysis of our VLBA
observations indicates that the maser emission does indeed arise along
the line of sight to a jet component in \m348, confirming this
prediction.  The very high linewidth occurring on such small spatial
scales and the rapid variability indicate that the H$_2$O emission is
more likely to arise from a shocked region at the interface between
the energetic jet material and the molecular gas in the cloud where
the jet is boring through, than simply as the result of amplification
by molecular clouds along the line of sight to the continuum jet.  The
orientation of the radio jets close to the plane of the sky also
results in shocks with the preferred orientation for strong masers
from our vantage point.  This hypothesis is supported by the spectral
evolution of the continuum source, which showed an inverted radio
spectrum with a peak at 22~GHz, later shifting to lower frequencies.
Further evidence for the ejection of new components in the jets is
shown by the sub-pc scale VLBA observations made over a period of 3
years.  In this scenario, the recent high frequency radio continuum
flare, the linear motion of the brightest continuum components and the
flare in the maser emission are all attributable to the generation and
ejection of new components in the approaching and receding jets.  The
very close temporal correlation between the flaring activity in the
maser emission and the continuum flare further suggest that the masing
region and the continuum hotspots are nearly equidistant from the
central engine and may be different manifestations of the same
dynamical events.

The gas in the molecular cloud within the central parsecs of \m348 has
a pre-shock density ranging from around 10$^6$ cm$^{-3}$ to a few
times 10$^9$ cm$^{-3}$.  An expanding bow-shock being driven into this
cloud by the AGN jet has a velocity between 135 \kms\ and 0.5c in the
direction of jet propagation, and between 35 \kms\ and 300 \kms\ at
various points along the oblique edges.  This shock generates a region
of very high temperature, ($\le$10$^5$ K), which dissociates the
molecular gas and to some extent shatters the dust grains expected to
be present and/or evaporates their icy mantles.  Immediately following
this shock, H$_2$ begins forming on the surviving dust grains when the
temperature has dropped to \ab1000 K, and this in turn provides
sufficient heating to stabilize the temperature at \ab400 K, with gas
densities of \ab10$^8$ cm$^{-3}$ and ultimately an \h2o\ abundance as
high as 10$^{-5}$.  Ammonia is not found because the levels of UV
radiation remain too high, but the detection of other molecular
species will help to test this scenario.  For example, we are
currently undertaking a search for SiO masers which might be generated
as the gas is liberated from the solid phase by the shattering of dust
grains toward the head of the shock.  We are also searching for
formaldehyde which might be located within the cocoon, and \HI\ from
the outer edges of the cloud which might be detectable in absorption
toward the background continuum source.

\begin{acknowledgements}

We thank Lincoln Greenhill, James Braatz, Mark Claussen, Greg Taylor,
Chris Carilli and Floris van der Tak for enlightening discussion
and/or comments. ABP is grateful to Barry Clark, Peggy Perley and the
data analysts at NRAO in Socorro for prompt scheduling of and
assistance with the VLBA experiment.  This research has made
 use of the NASA/IPAC Extragalactic Database (NED) which is
operated by the Jet Propulsion Laboratory, California Institute of
Technology, under contract with the National Aeronautics and Space
Administration.
\end{acknowledgements}
%\clearpage

\clearpage

% Table 1
% Version 1

\begin{deluxetable}{ccccccc}

%\rotate

%\small
%\voffset=-0.8in
%\hoffset=-0.5in
%\ptlandscape
%\large
%\pagestyle{empty}

\tablecolumns{7}
\tablewidth{0pc}
%\tablenum{1}
\tablecaption{VLBA Continuum Observations}
\tablehead{
\colhead{UT Date}                  &
\colhead{Frequency}         &
\colhead{Polarization}         &
\colhead{$B_{\rm maj}$}               &
\colhead{$B_{\rm min}$}               &
\colhead{PA}               &
\colhead{rms}              
\\
\colhead{                        }& 
\colhead{(GHz)                        }& 
\colhead{       }&
\colhead{(mas)} &
\colhead{(mas)} &
\colhead{(deg)} &
\colhead{(mJy beam$^{-1}$)} 
\\
%\colhead{ (1)}                    &
%\colhead{ (2)}                    &
%\colhead{ (3)}                    &
%\colhead{ (4)}                    &
%\colhead{ (5)}                    &
%\colhead{ (6)}                    &
%\colhead{ (7)}           
}
\startdata
1997.10&1.667&LCP&7.90&5.11&$-$24.5&0.06 \\
1997.10&4.987&LCP&2.66&1.67&$-$22.0&0.08 \\
1997.10&15.365&LCP&0.80&0.45&$-$15.0&0.54 \\
&&&&&&\\
1998.75&8.421&RCP&1.54&0.92&\ $-$1.5& 0.10 \\
1998.75&15.365&LCP&0.80&0.45&$-$15.0& 1.14 \\
1998.75&22.233&LCP&0.56&0.31&\ $-$6.9&0.63 \\
&&&&&&\\
2000.00&8.421&Dual&1.65&0.99&\ $-$3.0& 0.25 \\
2000.00&15.365&Dual&0.86&0.54&$-$11.1& 0.50 \\
2000.00&22.233&Dual&0.63&0.36&$-$10.4& 0.80 \\
\enddata
\label{tab:vlba-cont}
\end{deluxetable}

\clearpage

\begin{deluxetable}{ccccccc}
%\tabletypesize{\footnote}
%\rotate

%\small
%\voffset=-0.8in
%\hoffset=-0.5in
%\ptlandscape
%\large
%\pagestyle{empty}

\tablecolumns{7}
\tablewidth{0pc}
%\tablenum{1}
\tablecaption{Parsec-Scale Radio Properties}
\tablehead{
\colhead{Frequency}                  &
\colhead{Component}         &
\colhead{UT Date}         &
\colhead{$r$\tablenotemark{a}}              &
\colhead{PA}               &
\colhead{Flux Density}               &
\colhead{Size}              
\\
\colhead{(GHz)                        }& 
\colhead{                        }& 
\colhead{       }&
\colhead{(mas)} &
\colhead{(deg)} &
\colhead{(mJy)} &
\colhead{(mas)} 
\\
%\colhead{ (1)}                    &
%\colhead{ (2)}                    &
%\colhead{ (3)}                    &
%\colhead{ (4)}                    &
%\colhead{ (5)}                    &
%\colhead{ (6)}                    &
%\colhead{ (7)}           
}
\startdata
\underbar{1.7~GHz} 
&1+2+3&1997.10&0.0&\nodata&$30.6\pm 1.5$&$1.6\times 1.0$, PA $-8^\circ$ \\
&&&&&& \\
\underbar{5.0~GHz}
&1+2+3&1997.10&0.0&\nodata&$157.3\pm 8.8$& Unresolved \\
&&&&&& \\
\underbar{8.4~GHz}
&1&1998.75&0&\nodata&$228\pm 11$& Unresolved \\
&1A&2000.00&0&\nodata&$413\pm 20$& Unresolved \\
&1B&2000.00&0.311&-31&$423\pm 21$& Unresolved \\
&2&1998.75&1.526&-15&$27.9\pm 1.4$& Unresolved \\
&3&2000.00&2.519&-14&$35.4\pm 1.8$& Unresolved \\
&&&&&& \\
\underbar{15~GHz}
&1&1997.10&0&\nodata&$96\pm 5$&Unresolved \\
&1&1998.75&0&\nodata&$552\pm 28$&$0.16\times 0.11$, PA $-9^\circ$ \\
&1A&2000.00&0&\nodata&$537\pm 27$&Unresolved \\
&1B&2000.00&0.431&-15&$385\pm 19$&Unresolved \\
&2&1997.10&1.460&-16&$26\pm 2$&Unresolved \\
&2&1998.75&1.581&-15&$17\pm 1$&Unresolved \\
&2&2000.00&1.677&-12&$58\pm 3$&Unresolved \\
&3&2000.00&2.760&-12&$8\pm 1$&Unresolved \\
&&&&&& \\
\underbar{22~GHz}
&1&1998.75&0&\nodata&$530\pm 27$&$0.15\times 0.11$, PA $-22^\circ$ \\
&1A&2000.00&0&\nodata&$343\pm 17$&Unresolved \\
&1B&2000.00&0.432&-15&$362\pm 18$&Unresolved \\
&1A&2000.44&0&\nodata&$40\pm 2$&Unresolved \\
&C&2000.44&1.100&-16&$299\pm 15$&Unresolved \\
&1B&2000.44&1.473&-7.0&$175\pm 9$&Unresolved \\
&2&1998.75&1.599&-14&$16\pm 1$&Unresolved \\
&2&2000.00&1.587&-11&$46\pm 2$&$0.9\times 0.2$, PA $-9^\circ$ \\
&2&2000.44&1.960&-9.0&$115\pm 6$&Unresolved \\
\enddata
\tablenotetext{a}{with respect to southernmost component}
\label{tab:vlba-res}
\end{deluxetable}
%\footnotesize{$^\dag$ with respect to southernmost component}
\clearpage

%{\centerline{\bf Figure Captions}}

\end{document}